Rapid, high-sensitivity detection of biomolecules using dual-comb biosensing: application to the SARS-CoV-2 nucleocapsid protein

Shogo Miyamura[1,†], Ryo Oe[1,†], Takuya Nakahara[1], Shota Okada[2], Shuji Taue[3], Yu Tokizane[4], Takeo Minamikawa[5], Taka-aki Yano[4], Kunihiro Otsuka[5,6], Ayuko Sakane[5,7], Takuya Sasaki[5,7], Koji Yasutomo[5,6], Taira Kajisa[5,8,*], and Takeshi Yasui[4,*]

[1]Graduate School of Advanced Technology and Science, Tokushima University, 2-1 Minami-Josanjima, Tokushima, Tokushima 770-8506, Japan

[2]Graduate School of Sciences and Technology for Innovation, Tokushima University, 2-1 Minami-Josanjima, Tokushima, Tokushima 770-8506, Japan

[3]School of System Engineering, Kochi University of Technology, 185 Miyanokuchi, Tosayamada, Kami, Kochi 782-8502, Japan

[4]Division of Next-generation Photonics, Institute of Post-LED Photonics (pLED), Tokushima University, 2-1 Minami-Josanjima, Tokushima, Tokushima 770-8506, Japan

[5]Division of Interdisciplinary Researches for Medicine and Photonics, Institute of Post-LED Photonics (pLED), Tokushima University, 2-1 Minami-Josanjima, Tokushima, Tokushima 770-8506, Japan

[6]Department of Immunology and Parasitology, Graduate School of Medicine,

Tokushima University, 3-18-15 Kuramoto, Tokushima, 770-8503, Japan

[7]Department of Biochemistry, Graduate School of Medicine, Tokushima University, 3-18-15 Kuramoto, Tokushima, 770-8503, Japan

[8]Graduate School of Interdisciplinary New Science, Toyo University, 2100 Kujirai, Kawagoe, Saitama, 350-8585, Japan

[†]These authors contributed equally to this article.

*Correspondence:

Takeshi Yasui, yasui.takeshi@tokushima-u.ac.jp

Taira Kajisa, kajisa@toyo.jp

## Abstract

Rapid, sensitive detection of biomolecules is important for improved testing methods for viruses as well as biomarkers and environmental hormones. For example, testing for SARS-CoV-2 is essential in the fight against the COVID-19 pandemic. Reverse-transcription polymerase chain reaction (RT-PCR) is the current standard for COVID-19 testing; however, it is hampered by the long testing process. Shortening the testing process while achieving high sensitivity would facilitate sooner quarantine and thus presumably prevention of the spread of SARS-CoV-2. Here, we aim to achieve rapid, sensitive detection of the SARS-CoV-2 nucleocapsid protein by enhancing the performance of optical biosensing with a dual-comb configuration of optical frequency combs. The virus-concentration-dependent optical spectrum shift is transformed into a photonic RF shift by frequency conversion between the optical and RF regions, facilitating mature electrical frequency measurements. Furthermore, active-dummy temperature-drift compensation enables very small changes in the virus-concentration-dependent signal to be extracted from the large, variable background signal. This dual-comb biosensing technique has the potential to reduce the COVID-19 testing time to 10 min while maintaining sensitivity close to that of RT-PCR. Furthermore, this system can be applied for sensing of not only viruses but also various biomolecules for clinical diagnosis, health care, and environmental monitoring.

Biosensors are biomolecular sensors that utilize the skilful molecular identification function of living organisms; they are applied to a wide range of fields, such as medical care, food industry, and environmental monitoring. However, further enhancement of the biosensing performance is still required for improved testing methods for infectious pathogens as well as biomarkers and RNA. One timely and urgent application that benefits from improved performance is testing for coronavirus disease 2019 (COVID-19). COVID-19, caused by severe acute respiratory syndrome coronavirus 2 (SARS-CoV-2), has rapidly spread and is still occurring all over the world. One reason for the failure to suppress the rapid spread of COVID-19 is the time-consuming testing process for SARS-CoV-2 because it hinders sooner quarantine and thus presumably prevention of the spread of COVID-19. The current standard for COVID-19 testing is reverse-transcription polymerase chain reaction (RT-PCR) [1-3], which sensitively detects SARS-CoV-2 RNA within a limit of detection (LOD) range from sub-aM to several aM. However, RT-PCR requires multiple time-consuming steps (required time = 4~5 hours). Although a qualitative antigen test enables rapid, simple testing (required time = 15~30 min), it requires a certain amount of virus due to the limited sensitivity. To facilitate earlier quarantine, there is a considerable need for rapid, high-sensitivity detection of SARS-CoV-2.

Among the potential methods that may improve the SARS-CoV-2 detection time and sensitivity is the use of optical biosensors due to both their rapidity and high sensitivity [4,5]. For example, optical biosensors based on surface plasmon resonance (SPR) [6,7] have been widely used for analysing viruses [8-14], proteins [15,16], DNA [17,18], and whole cells [19,20]; the sample-concentration-dependent optical spectral shift of the SPR trough in the wavelength or angular spectrum is measured through the combined effect of SPR and the molecular identification function on the sensor surface. SPR enables real-time, label-free analysis of intermolecular interactions. The LOD for SARS-CoV-2 has reached 85 fM [14]; however, it is not yet at the RT-PCR level. A reason for the limited sensitivity is the optical instrumentation resolution as well as the relatively broad spectrum of the SPR trough compared to its slight spectrum shift.

One promising method to overcome the limitation of optical instrumentation resolution is to transform the sample-concentration-dependent optical spectral shift into its equivalent photonic radio-frequency (RF) spectral shift because such photonic RF biosensing would benefit from the high precision and real-time nature provided by mature electrical frequency measurements with frequency standards. Recently, optical frequency combs (OFCs) [21-24] have attracted attention for use as photonic RF sensors based on a frequency conversion function between the optical and RF regions [25-28]. An OFC is composed of a series of optical frequency modes (freq. = $v_m$) with a constant mode spacing of $f_{rep}$ in the RF band. The relation between $v_m$ and $f_{rep}$ is given by

$$v_m = f_{ceo} + m f_{rep}, \qquad (1)$$

where $f_{ceo}$ is the carrier-envelope-offset frequency and $m$ is the mode number. Thus,

the OFC acts as an accurate frequency converter between $\nu_m$ and $f_{rep}$. For example, a refractive-index (RI)-dependent optical spectrum shift was converted into a change in $f_{rep}$ by placing a multimode-interference (MMI) fibre sensor [27,28] inside a fibre OFC cavity to realize an RI-dependent variable-optical-bandpass filter [29-31]. Then, the $f_{rep}$ signal was rapidly and precisely measured by an RF frequency counter. The photonic-to-RF conversion reduces the spectral linewidth to below 1 Hz [29]. Furthermore, the intracavity fibre sensor enables multiple interactions between the sample and the light, reducing the resolution to $4.9 \times 10^{-6}$ refractive index units (RIU), which is two orders of magnitude better than that in a previous study [27]. Such RI-sensing OFCs would have the potential to be further extended to optical biosensing through surface modification of the MMI fibre sensor with a molecular identification layer in terms of biomolecule interactions, similar to SPR. However, there are no attempts to apply such RI-sensing OFCs for optical biosensing because the residual temperature drift of $f_{rep}$ (typically, a few hundreds of Hz/hour) is larger than the sample-concentration-dependent $f_{rep}$ shift (typically, a few to a few tens of Hz), hindering their extension to biosensing OFCs. Largely reducing the temperature drift of $f_{rep}$ is essential to achieve both a sensitivity close to that of RT-PCR and a measurement time considerably shorter than that of RT-PCR.

    Thus, in this article, we first developed a dual-comb configuration [32] with an active sensing OFC and a dummy OFC to suppress the temperature drift of $f_{rep}$, namely, dual-comb biosensing; this function is similar to the active-dummy temperature compensation of strain sensors. Then, for a preliminary test, we applied the active-dummy dual-sensing OFCs to RI sensing of a glycerol solution. Finally, as a proof of concept, we demonstrated rapid detection of the SARS-CoV-2 nucleocapsid protein (N protein) antigen by combining dual-comb biosensing and surface modification with the SARS-CoV-2 N protein antibody.

## Results

**General principle of operation**

In this study, we sought to design a biosensor that combines photonic-to-RF conversion and antigen-antibody interactions in an OFC. The biosensing OFC operates through three steps: (1) antigen-antibody interactions on the antibody-modified sensor surface, (2) RI-dependent optical spectrum shift of the OFC provided by the intracavity MMI fibre sensor [27,28,33], and (3) photonic-to-RF conversion via the wavelength dispersion of the fibre cavity [29-31], as depicted in Fig. 1a. The functions of steps (1) and (2) are implemented by an intracavity MMI fibre sensor with antibody surface modification, as shown in Fig. 1b. Finally, the change in the antigen concentration can be read out as the $f_{rep}$ shift via $f_{rep} = c/nL$, where $c$ is the velocity of light in vacuum and $nL$ is the optical cavity length. The $f_{rep}$ linewidth is much smaller than the $f_{rep}$ shift expected due to antigen concentration changes.

**Temperature drift in the single-comb configuration**

We first evaluated the cavity temperature dependence of $f_{rep}$ in a single sensing OFC because temperature disturbances cause $f_{rep}$ to fluctuate via thermal changes of $nL$. To this end, we measured the temporal fluctuations in the $f_{rep}$ of the single sensing OFC under an uncontrolled cavity temperature, as shown in Fig. 2a. Pure water was used as a standard sample with a stable RI and placed in a glass sample cell together with the MMI fibre sensor without surface modification for RI sensing. The output light from the OFC was detected by a photodetector (PD), and $f_{rep}$ was measured by an RF frequency counter synchronized to a rubidium frequency standard working in the RF band. Figure 2b shows the $f_{rep}$ shift ($\delta f_{rep}$) when the cavity temperature changed over a range of 1 °C. $\delta f_{rep}$ represents the frequency deviation from the initial measurement value. The temporal behaviour of $\delta f_{rep}$ in synchronization with the cavity temperature indicated a temperature sensitivity of approximately -400 Hz/°C. This cavity-temperature-dependent $f_{rep}$ drift is considerably larger than the sample-concentration-dependent $f_{rep}$ shift in biosensing (typically, a few to a few tens of Hz). Although the cavity temperature was actively controlled within a range of 25.0±0.1 °C in the following experiments, this was still insufficient to suppress the cavity-temperature-dependent $f_{rep}$ drift to below the sample-concentration-dependent $f_{rep}$ shift. Thus, to further reduce the temperature drift, we applied a dual-comb configuration for active-dummy temperature compensation together with active control of the cavity temperature, as described in the following subsection.

**Active-dummy compensation of the temperature drift with the dual-comb configuration**

A dual-comb configuration [32] with an active sensing OFC with a frequency spacing of $f_{rep1}$ and a dummy sensing OFC with a frequency spacing of $f_{rep2}$ was adopted to compensate for the temperature drift. Figure 3a shows a schematic drawing of the dual-comb configuration, in which a pair of fibre OFC cavities were arranged in a temperature-controlled box so that they were affected by similar temperature fluctuations. In this configuration, although $f_{rep1}$ and $f_{rep2}$ fluctuate depending on the residual fluctuation of the cavity temperature, their fluctuations are similar because they experience the same thermal disturbances. Therefore, the frequency difference $\Delta f_{rep}$ between $f_{rep1}$ and $f_{rep2}$ remains constant regardless of the temperature drift of $f_{rep1}$ and $f_{rep2}$. Thus, when the active sensing OFC evaluates a sample solution in a certain temperature environment and the dummy sensing OFC evaluates a reference material in the same temperature environment, $\Delta f_{rep}$ reflects the sample concentration without influence from temperature drift. Figures 3b, 3c, and 3d show the MMI and sample cell for dual-comb RI sensing of pure water, dual-comb RI sensing of glycerol solution, and dual-comb biosensing of the SARS-CoV-2 N protein antigen, respectively. Table 1 summarizes $f_{rep1}$, $f_{rep2}$, $\Delta f_{rep}$, the MMI, and the sample cell used in the following three dual-comb sensing experiments; these frequency values were selected for stable operation and better temperature compensation. A pair of output light beams from the active and dummy sensing OFCs was detected by a pair of PDs; then, $f_{rep1}$, $f_{rep2}$, and $\Delta f_{rep}$ were measured by the RF frequency counter (see the Methods section).

The blue and green lines in Fig. 4 show the temporal shifts in $f_{rep1}$ and $f_{rep2}$, namely, $\delta f_{rep1}$ and $\delta f_{rep2}$, respectively, when pure water was used as a sample for both the active and dummy sensing OFCs without surface modification (see Table 1 and Fig. 3b). $\delta f_{rep1}$ and $\delta f_{rep2}$ suffered from a temperature drift of over -38 Hz due to an increase in the cavity temperature; however, they behaved almost the same in terms of drift. The resulting $\Delta f_{rep}$ shift ($\delta\Delta f_{rep}$) was stable, with a variation of up to 1.18 Hz, as shown by the red line in Fig. 4. This level of $\Delta f_{rep}$ stability is sufficient for precise measurement of the sample-concentration-dependent $\Delta f_{rep}$ shift.

We next tested active-dummy temperature compensation for RI sensing of a liquid sample different from the reference sample. For RI sensing, the active and dummy sensing OFCs have no surface modification (see Table 1 and Fig. 3c). We used glycerol solutions consisting of glycerine and pure water at different ratios, corresponding to different RIs, as target samples in the active sensing OFC. We exchanged the target sample and the reference sample (pure water) by using a pair of peristaltic pumps. The blue and green lines in Fig. 5a represent $\delta f_{rep1}$ and $\delta f_{rep2}$ as the glycerol solution concentration increased from 0 vol% to 5 vol%. $\delta f_{rep2}$, in the dummy sensing OFC, exhibited a slow drift with some rapid fluctuations. Since these rapid fluctuations were synchronized with the operation of the peristaltic pump, they were caused by disturbances from the water flow when the samples were exchanged. In contrast, $\delta f_{rep1}$, in the active sensing OFC, exhibited a combination of a step-like change with the sample RI and the slow drift shown by $\delta f_{rep2}$. This combined behaviour of $\delta f_{rep1}$ is detrimental to the RI sensing performance in the single sensing OFC configuration.

Figure 5b shows a sensorgram of $\delta\Delta f_{rep}$ calculated by subtracting the green line ($\delta f_{rep2}$) from the blue line ($\delta f_{rep1}$) in Fig. 5a. The temperature drift almost disappeared, and only the step-like change with the sample RI was present in $\delta\Delta f_{rep}$. From the mean of $\delta\Delta f_{rep}$ at each concentration, we calculated a relationship between the sample RI and $\delta\Delta f_{rep}$, indicated by the red line in Fig. 5c. The good fitting result indicated that the dual-comb effect minimizes the effect of temperature fluctuations.

**Rapid detection of the SARS-CoV-2 N protein antigen**

Antibody modification of the intracavity MMI fibre sensor creates a photonic RF biosensor for the detection of target antigens through antibody-antigen reactions (see Fig. 1a). The enhanced RI precision above covers the RI change expected for antigen-antibody interactions, enabling us to apply these dual-sensing OFCs to rapid, high-sensitivity detection of viruses/pathogens and biological molecules.

The concept of antigen-antibody interactions (in this case, a viral protein) was applied for detection of SARS-CoV-2 protein. Among several proteins in SARS-CoV-2, the N protein, which functions to package the viral RNA genome within the viral envelope into a ribonucleoprotein complex, is a promising candidate for antigen-antibody interactions because of its abundance, low probability of mutation, and relatively low molecular weight. Thus, we used the combination of a commercialized N protein monoclonal antibody and a commercialized N protein recombinant antigen, which exhibited high affinity in an enzyme-linked immunosorbent assay, for the intracavity MMI biosensor (see Fig. 1b). The MMI fibre sensors with and without surface modification by the immobilized antibody were placed together in the same sample cell as the active and dummy sensing OFCs, respectively (see Table 1 and Fig. 3d). Solution samples of the N protein antigen in phosphate-buffered saline (PBS) at different molar concentrations were consecutively introduced into the sample cell with a peristaltic pump. The N protein antigen-antibody interaction could only occur on the sensor surface of the active sensing OFC because the surface of the dummy sensing OFC did not include immobilized antibodies. Thus, the dummy sensing OFC was used as a negative control.

Figure 6a shows the sensorgram of $\delta f_{rep1}$ and $\delta f_{rep2}$ as the molar concentration of the antigen/PBS solution increased from 1 aM ~ 1 nM after starting with pure PBS; this range of molar concentrations was selected for the sample considering the LOD for SARS-CoV-2 of RT-PCR. The time period for data acquisition (see colour-highlighted zones in Fig. 6a) was set to 10 min. Each grey zone (time period = 8 min) includes the sample change by the pump, the waiting time for completion of the antigen-antibody interaction, and rinsing of the sensor surface with PBS. Since the step-like change in $\delta f_{rep1}$ with the antigen concentration was completely overshadowed by the background temperature drift, we calculated the frequency difference ($\delta\Delta f_{rep}$) between $\delta f_{rep1}$ and $\delta f_{rep2}$ to eliminate the influence of temperature drift as described in the previous subsection. Figure 6b shows the sensorgram of $\delta\Delta f_{rep}$.

Focusing on the zones highlighted in colours other than grey, a slightly dull stepped change in $\delta\Delta f_{rep}$ dependent on the molar concentration was observed, although a small drift in $\delta\Delta f_{rep}$ within the range of a few Hz remained at each molar concentration. To evaluate the validity of this behaviour, we calculated the relationship between the molar concentration and $\delta\Delta f_{rep}$, as shown by the red circles plotted in Fig. 6c. The negative slope was consistent with the RI dependence of $\delta\Delta f_{rep}$ (see the red line plotted in Fig. 5c) because the progression of the antigen-antibody reaction increases the effective RI near the MMI fibre sensor and hence decreases $f_{rep}$ [34]. In this way, we demonstrated the potential for rapid detection of the SARS-CoV-2 N protein antigen within this range of molar concentrations.

**Discussion**

To achieve rapid, high-sensitivity biosensing, we developed dual-comb biosensing and applied it to the detection of the SARS-CoV-2 N protein antigen with a molar concentration from 1 aM to 1 µM in a measurement time of 10 min. We performed the quantitative analysis of the result shown in Fig. 6c. The antigen-antibody reaction is represented by a sigmoidal curve, often used when discussing the biosenser performance [35]; thus, a sigmoidal curve was applied to the experimental data to evaluate the ability of dual-comb biosensing to sense the SARS-CoV-2 N protein. The sigmoidal function of the Hill plot is given by

$$\delta\Delta f_{rep} = \delta\Delta f_{rep\_min} + \frac{\delta\Delta f_{rep\_max} - \delta\Delta f_{rep\_min}}{1 + \left(\frac{[C]}{[C_{Ka}]}\right)^n}, \tag{2}$$

where $\delta\Delta f_{rep\_max}$ and $\delta\Delta f_{rep\_min}$ are the maximum and minimum of $\delta\Delta f_{rep}$, $n$ is the Hill coefficient, $C$ is the concentration of the antigen, and $C_{Ka}$ is the dissociation constant. The purple line in Fig. 6c represents the sigmoidal fit. At the end of the grey zones in Fig. 6b, the $\delta\Delta f_{rep}$ signal reflects the amount of antigen adsorbed on the sensor surface after desorption. From the curve fitting analysis, we determined $C_{Ka}$ to be 8.4 fM, indicating the molar concentration at the middle between $\delta\Delta f_{rep\_max}$ and $\delta\Delta f_{rep\_min}$ (see pink dashed lines in Fig. 6c). When the linear range (LR) was defined as the molar concentration at 10~90 % of the dynamic signal range of $\delta\Delta f_{rep\_max}$ to $\delta\Delta f_{rep\_min}$ (see blue dashed lines in Fig. 6c) [35], it was determined to be 34 aM ~ 2.1 pM. Additionally, the LOD was calculated to be 38 aM from the crossover point between $\delta\Delta f_{rep\_max}$ and the linear approximation given by the Hill coefficient $n$ (= 0.40, see the green lines in Fig. 6c).

We next compare the dual-comb biosensing with other SARS-CoV-2 testing methods, as shown in Table 2. RT-PCR testing shows an LOD of ~ 100 copies/ml [36], corresponding to 0.17 aM; however, it requires a long analysis time. SPR [37] and its enhancement [14,38] could achieve an LOD of sub-pM ~ pM in rapid analysis. The LOD of the colorimetric assay remains at approximately a few tens of pM [39]. Importantly, only dual-comb biosensing achieves an LOD close to that of RT-PCR with rapid measurement considerably shorter than that of RT-PCR.

We also discuss the potential for dual-comb biosensing to be used for the detection of other viruses and biomolecules of interest. The considerably low $C_{Ka}$ implies the potential for a wide variety of biosensing applications, for example, early detection of cancer cells from a droplet of blood or other body fluids by detecting a sugar chain specifically expressed on the cell surface. Such blood biopsy or liquid biopsy will be a powerful tool for detection of important biomarkers, such as proteins or RNA, in addition to cancer cells. As another interesting application, biosensing of exosomes via miRNA is expected to make a great contribution to the diagnosis (marker) and treatment (drug delivery) of diseases such as cancer and Alzheimer's disease because exosomes play an important role as an intercellular communication tool.

## Conclusion

We have demonstrated dual-comb biosensing for rapid, high-sensitivity detection of biomolecules. To the best of our knowledge, this is the first application of an OFC as a biosensor itself. The integration of photonic-to-RF conversion, an intracavity sensor, and active-dummy dual-comb compensation in the OFC enables detection of the SARS-CoV-2 N protein antigen with an LOD of 38 aM, close to that of RT-PCR, in a measurement time of 10 min, considerably shorter than that of RT-PCR. The current COVID-19 pandemic may diminish in the future; however, we are always at risk of facing another emerging and re-emerging infectious disease again. As dual-comb biosensing can be used for other viruses through selection of the antigen-antibody reaction or other molecular identifications, it will be important as a proactive measure against unknown infectious diseases. Simultaneous achievement of high sensitivity and rapid measurement by dual-comb biosensing will greatly enhance the applicability of biosensors to viruses, biomarkers, environmental hormones, and so on.

**Methods**

**MMI fibre sensor**

Figure 1b shows a schematic diagram of the intracavity MMI fibre sensor with antibody surface modification. When the surface of the fibre sensor is not modified with an antibody, the MMI fibre sensor functions as an RI sensor. The MMI fibre sensor is composed of a clad-less multimode fibre (MMF; Thorlabs Inc., Newton, NJ, USA, FG125LA, core diameter= 125 µm, fibre length = 58.94 mm) with a pair of single-mode fibres (SMFs) at both ends (Corning Inc., Corning, NY, USA, SMF28e+, core diameter = 8.2 µm, cladding diameter=125 µm, fibre length = 150 mm) [29-31]. Only the exposed core of the clad-less MMF functions as a sensing part. The OFC light passing through the input SMF is diffracted at the entrance face of the clad-less MMF and then undergoes repeated total internal reflection at the boundary between the clad-less MMF core surface and the sample solution. Only the OFC modes satisfying the MMI wavelength $\lambda_{MMI}$ can exit through the clad-less MMF and then be transmitted through the output SMF. $\lambda_{MMI}$ is given by

$$\lambda_{MMI} = \frac{n_{MMF} m_{MMI}}{L_{MMF}} \left[ D(n_{sam}) \right]^2, \qquad (3)$$

where $L_{MMF}$ and $n_{MMF}$ are the geometrical length and RI of the clad-less MMF, $m_{MMI}$ is the order of the MMI, $n_{sam}$ is the RI near the clad-less MMF core surface (namely, sample RI), and $D(n_{sam})$ is the effective core diameter of the clad-less MMF. Since $D(n_{sam})$ is influenced by the Goos-Hänchen shift on the core surface of the clad-less MMF, $\lambda_{MMI}$ is a function of the sample RI near the sensor surface. The intracavity MMI fibre sensor in this study functions as an RI-dependent optical bandpass filter tuneable around $\lambda_{MMI}$ (= 1556.6 nm) with constructive interference at $m$ = 4. This $\lambda_{MMI}$ was selected to match a spectral peak of the fibre OFC, suppressing the power loss. The RI-dependent $\lambda_{MMI}$ shift of the OFC is converted into an RI-dependent $f_{rep}$ shift via the wavelength dispersion of the cavity fibre (see Fig. 1a) [29]. Furthermore, if the surface of the MMI fibre sensor is modified with a virus antibody, then the RI-dependent $f_{rep}$ shift is converted into a virus-antigen-concentration-dependent $f_{rep}$ shift through antibody-antigen reactions. In other words, the intracavity MMI fibre sensor with antibody surface modification enables a photonic RF biosensor for viruses.

**Single-comb configuration of the sensing OFC**

We used a linear fibre cavity mode-locked by a saturable absorber mirror for easy mode-locked oscillation and compact size (see Fig. 2a). The linear cavity includes a 2.6-m-long SMF (Corning Inc., Corning, NY, USA, SMF28e+, dispersion at 1550 nm = 17 ps•km$^{-1}$•nm$^{-1}$), a 0.6-m-long erbium-doped fibre (EDF; nLIGHT Inc., Camas, WA, USA, LIEKKI ER30-4/125, dispersion at 1550 nm = -22.75 ps•km$^{-1}$•nm$^{-1}$), a saturable absorber mirror (BATOP GmbH, Jena, Germany, SAM-1550-55-2ps-1.3b-0, high reflection band = 1480-1640 nm, absorbance = 55 %, modulation depth = 2.4 %, relaxation time constant = ~2 ps, size = 1.3 mm width, 1.3 mm height, 0.4 mm thickness), a wavelength-division-multiplexing coupler (WDM; AFR Ltd., Zhuhai, China, WDM-1-9855-N-B-1-F), a pumping laser diode (LD pump source; Thorlabs Inc., Newton, NJ, USA, BL976-PAG700, wavelength = 976 nm, power = 700 mW), a 90:10 fibre output coupler (OC; AFR Ltd., Zhuhai, China, PMOFM-55-2-B-Q-F-90), and an intracavity MMI fibre sensor (MMI). The total dispersion of the fibre cavity was set to -0.12 pm/s$^2$ for stable operation. The fibre cavity was placed in an aluminium box without a lid, and its temperature was not actively controlled. We set the frequency spacing of the sensing OFC to approximately 31.7 MHz for stable mode-locked oscillation with the intracavity MMI fibre sensor. The light output of the sensing OFC was detected by a photodetector (PD; Thorlabs Inc., Newton, NJ, USA, PDA05CF2, wavelength = 800~1700 nm, frequency bandwidth = 150 MHz), and the resulting frequency signal of $f_{rep}$ was measured by an RF frequency counter (Keysight Technologies, Santa Rosa, CA, USA, 53230A, frequency resolution = 12 digit•s$^{-1}$) synchronized to a rubidium frequency standard (Stanford Research Systems Inc., Sunnyvale, CA, USA, FS725, accuracy = 5×10$^{-11}$ and instability = 2×10$^{-11}$ at 1 s).

**Dual-comb configuration of active and dummy sensing OFCs**

We used a pair of linear-cavity sensing OFCs (frequency spacing = $f_{rep1}$ and $f_{rep2}$, frequency difference between them = $\Delta f_{rep} = f_{rep1} - f_{rep2}$) for the active sensing OFC and the dummy sensing OFC in the dual-comb configuration (see Fig. 3a). The configuration of each linear cavity was similar to that in Fig. 2a. The output light of the LD pump source was split into two beams and used for these two OFCs, which eliminates the influence of power fluctuations in the LD pump source through common-mode rejection. These OFC fibre cavities were enclosed in an aluminium box with a lid, and its temperature was controlled to 25.0±0.1 °C by a combination of a Peltier heater (Kaito Denshi Inc., Hasuda, Saitama, Japan, TEC1-12708, power = 76 W), a thermistor (YHTC Co., Ltd., Machida, Tokyo, Japan, PB7-42H-K1), and a temperature controller (Thorlabs Inc., Newton, NJ, USA, TED200C, PID control) (not shown in Fig. 3a). We set the frequency spacings of the active and dummy sensing OFCs to approximately 31.7~32.5 MHz in RI sensing and 29.6~29.7 MHz in biosensing by adjusting the cavity length; the resulting $\Delta f_{rep}$ was -851.9 kHz in RI sensing and -88.6 kHz in biosensing. We set the cavity fibres of the dual OFCs in the same arrangement so that they would be equally affected by environmental temperature disturbances [32]. The light output of the dual OFCs was detected by a pair of PDs, and the resulting frequency signals of $f_{rep1}$, $f_{rep2}$ and $\Delta f_{rep}$ were measured through a combination of an RF frequency counter and a rubidium frequency standard.

**Antibody modification of the MMI fibre sensor**

A schematic diagram of the intracavity MMI fibre sensor with antibody surface modification is shown in Fig. 1b. First, a UV ozone cleaner (Sun Energy Corp., Minoo, Osaka, Japan, SKB1101N-01) was applied to the MMI fibre sensor for 30 min to remove any organic compounds on the surface of the clad-less MMF and modify the resulting surface with hydroxy groups. Second, the surface of the clad-less MMF was modified with amino-terminated groups through a silane coupling reaction using 1 % (v/v) 3-aminopropyltriethoxysilane (APTS) in ethanol for 1 hour, followed by washing with ultrapure water and drying at 110 °C for 10 min. Third, the monoclonal antibody specific for the N protein antigen was immobilized on the amino-group-coated MMF by a dehydration-condensation reaction using 10 mM 4-(4,6-dimethoxy-1,3,5-triazin-2-yl)-4-methylmorpholinium chloride (DMTMM) in PBS buffer (pH 7.4).

**Data analysis**

The frequency spacings of the OFCs ($f_{rep}$, $f_{rep1}$, and $f_{rep2}$) were continuously acquired by an RF frequency counter with a gate time of 100 ms and a sampling interval of 2.8 s. The frequency difference $\Delta f_{rep}$ between $f_{rep1}$ and $f_{rep2}$ was calculated from acquired $f_{rep1}$ and $f_{rep2}$. Finally, $\delta f_{rep}$, $\delta f_{rep1}$, $\delta f_{rep2}$, and $\delta\Delta f_{rep}$ were calculated as the frequency deviations from the initial values of $f_{rep}$, $f_{rep1}$, $f_{rep2}$, and $\Delta f_{rep}$, respectively. In the dual-comb biosensing of SARS-CoV-2 N protein antigen, we calculated the 99.9 % confidence interval for the first 100 data of the $\delta\Delta f_{rep}$ sensorgram measured in the PBS, and then used it as a criterion of rejection test to judge whether $\delta\Delta f_{rep}$ value acquired at each molar concentration is considered as a measurement error.

**Data availability**

The data that support the findings of this study are available from the corresponding author upon reasonable request.

**References**

[1] Suo, T., Liu, X., Feng, J., Guo, M., Hu, W., Guo, D., Ullah, H., Yang, Y., Zhang, Q., Wang, X., Sajid, M., Huang, Z., Deng, L., Chen, T., Liu, F., Xu, K., Liu, Y., Zhang, Q., Liu, Y., Xiong, Y., Chen, G., Lan, K. & Chen, Yu. ddPCR: a more accurate tool for SARS-CoV-2 detection in low viral load specimens. *Emerg. Microbes & infec.* **9**, 1259-1268 (2020).

[2] Panpradist, N., Qin, W., Ruth, P. S., Kotnik, J. H., Oreskovic, A. K., Miller, A., Stewart, S. W.A., Vrana, J., Han, P. D., Beck, I. A., Starita, L. M., Frenkel, L. M. & Barry R. Lutz. Simpler and faster Covid-19 testing: Strategies to streamline SARS-CoV-2 molecular assays. *EBioMedicine* **64**, 103236 (2021).

[3] Oranger, A., Manzari, C., Chiara, M., Notario, E., Fosso, B., Parisi, A., Bianco, A., Iacobellis, M., d'Avenia, M., D'Erchia, A. M. & Pesole, G. Accurate detection and quantification of SARS-CoV-2 genomic and subgenomic mRNAs by ddPCR and meta-transcriptomics analysis. *Commun. Biol.* **4**, 1215 (2021).

[4]     Borisov, S. M. & Wolfbeis, O. S. Optical biosensors. *Chem. Rev.* **108**, 423-461 (2008).

[5]     Damborský, P., Švitel, J. & Katrlík, J. Optical biosensors. *Essays Biochem.* **60**, 91-100 (2016).

[6]     Homola, J., Yee, S. S. & Gauglitz, G. Surface plasmon resonance sensors. *Sens. Actuators B* **54**, 3–15 (1999).

[7]     Pattnaik, P. Surface plasmon resonance. *Appl. Biochem. Biotechnol.* **126**, 079–092 (2005).

[8]     Liu, Y. & Huang, C. Z. One-step conjugation chemistry of DNA with highly scattered silver nanoparticles for sandwich detection of DNA. *Analyst* **137**, 3434-3436 (2012).

[9]     Omar, N. A. S., Fen, Y. W., Abdullah, J., Sadrolhosseini, A. R., Mustapha Kamil, Y., Fauzi, N. I. M., Hashim, H. S. & Mahdi, M. A. Quantitative and selective surface plasmon resonance response based on a reduced graphene oxide-polyamidoamine nanocomposite for detection of dengue virus e-proteins. *Nanomaterials* **10**, 569 (2020).

[10]    Ashiba, H., Sugiyama, Y., Wang, X., Shirato, H., Higo-Moriguchi, K., Taniguchi, K., Ohki, Y. & Fujimaki, M. Detection of norovirus virus-like particles using a surface plasmon resonance-assisted fluoroimmunosensor optimized for quantum dot fluorescent labels. *Biosens. Bioelectron.* **93**, 260-266 (2017).

[11]    Bai, H., Wang, R., Hargis, B., Lu, H. & Li, Y. A SPR aptasensor for detection of

avian influenza virus H5N1. *Sensors* **12**, 12506-12518 (2012).

[12]   Djaileb, A., Charron, B., Jodaylami, M. H., Thibault, V., Coutu, J., Stevenson, K., Forest, S., Live, L. S., Boudreau, D., Pelletier, J. N. & Masson, J. F. A rapid and quantitative serum test for SARS-CoV-2 antibodies with portable surface plasmon resonance sensing (2020).

[13]   Moznuzzaman, M., Khan, I. & Islam, M. R. Nano-layered surface plasmon resonance-based highly sensitive biosensor for virus detection: A theoretical approach to detect SARS-CoV-2. *AIP Adv.* **11**, 065023 (2021).

[14]   Yano, T. A., Kajisa, T., Ono, M., Miyasaka, Y., Hasegawa, Y., Saito, A., Otsuka, K., Sakane, A., Sasaki, T., Yasutomo, K., Hamajima, R., Kanai, Y., Kobayashi, T., Matsuura, Y., Itonaga, M. & Yasui, T. Ultrasensitive detection of SARS-CoV-2 nucleocapsid protein using large gold nanoparticle-enhanced surface plasmon resonance. *Sci. Rep.* **12**, 1-8 (2022).

[15]   Bian, S., Lu J., Delport, F., Vermeire, S., Spasic, D., Lammertyn, J. & Gils, A. Development and validation of an optical biosensor for rapid monitoring of adalimumab in serum of patients with Crohn's disease. *Drug Test. Anal.* **10**, 592-596 (2018).

[16]   Singh, M., Holzinger, M., Tabrizian, M., Winters, S., Berner, N. C., Cosnier, S. & Duesberg, G. S. Noncovalently functionalized monolayer graphene for sensitivity enhancement of surface plasmon resonance immunosensors. *J. Am. Chem. Soc.* **137**, 2800-2803 (2015).

[17]   Pollet, J., Janssen, K, P. F., Knez, K., Lammertyn, J. Real-time monitoring of solid-phase PCR using fiber-optic SPR. *Small* **7**, 1003–1006 (2011).

[18]   Daems, D., Knez, K., Delport, F., Spasic, D. & Lammertyn, J. Real-time PCR melting analysis with fiber optic SPR enables multiplex DNA identification of bacteria. *Analyst* **141**, 1906-1911 (2016).

[19]   Malachovská, V. M., Ribaut, C., Voisin, V. V., Surin, M., Leclè, P., Wattiez, R. & Caucheteur, C. Fiber-optic SPR immunosensors tailored to target epithelial cells through membrane receptors. *Anal. Chem.* **87**, 5957-5965 (2015).

[20]   Leonard, P., Hearty, S., Quinn, J. & O'Kennedy, R. A generic approach for the detection of whole Listeria monocytogenes cells in contaminated samples using surface plasmon resonance. *Biosens. Bioelectron.* **19**, 1331-1335 (2004).

[21]   Udem, T., Reichert, J., Holzwarth, R. & Hänsch, T. W. Accurate measurement of large optical frequency differences with a mode-locked laser. Opt. Lett. 24, 881–883 (1999).

[22]   Niering, M., Holzwarth, R., Reichert, J., Pokasov, P., Udem, T., Weitz, M., Hänsch, T. W., Lemonde, P., Santarelli, G., Abgrall, M., Laurent, P., Salomon, C. & Clairon, A. Measurement of the hydrogen 1S-2S transition frequency by phase coherent comparison with a microwave cesium fountain clock. *Phys. Rev. Lett.* **84**, 5496–5499 (2000).

[23]   Jones, D. J., Diddams, S. A., Ranka, J. K., Stentz, A., Windeler, R. S., Hall, J. L. & Cundiff, S. T. Carrier-envelope phase control of femtosecond mode-locked

lasers and direct optical frequency synthesis. *Science* **288**, 635–639 (2000).

[24] Udem, T., Holzwarth, R. & Hänsch, T. W. Optical frequency metrology. *Nature* **416**, 233–237 (2002).

[25] Wang, S., Lu, P., Liao, H., Zhang, L., Liu, D. & Zhang, J. Passively mode-locked fiber laser sensor for acoustic pressure sensing. *J. Mod. Opt.* **60**, 1892-1897 (2013).

[26] Minamikawa, T., Ogura, T., Nakajima, Y., Hase, E., Mizutani, Y., Yamamoto, H., Minoshima, K. & Yasui, T. Strain sensing based on strain to radio-frequency conversion of optical frequency comb. *Opt. Express* **26**, 9484-9491 (2018).

[27] Taue, S., Matsumoto, Y., Fukano, H. & Tsuruta, K. Experimental analysis of optical fiber multimode interference structure and its application to refractive index measurement. *Jpn. J. Appl. Phys.* **51**, 04DG14 (2012).

[28] Fukano, H., Watanabe, D. & Taue, S. Sensitivity characteristics of multimode-interference optical-fiber temperature-sensor with solid cladding material. *IEEE Sens. J.* **16**, 8921-8927 (2016).

[29] Oe, R., Taue, S., Minamikawa, T., Nagai, K., Shibuya, K., Mizuno, T., Yamagiwa, M., Mizutani, Y., Yamamoto, H., Iwata, T., Fukano, H., Nakajima, Y., Minoshima, K. & Yasui, T. Refractive-index-sensing optical comb based on photonic radio-frequency conversion with intracavity multi-mode interference fiber sensor. *Opt. Express* **26**, 19694-19706 (2018).

[30] Oe, R., Minamikawa, T., Taue, S., Koresawa, H., Mizuno, T., Yamagiwa, M.,

Mizutani, Y., Yamamoto, H., Iwata, T. & Yasui, T. Refractive index sensing with temperature compensation by a multimode-interference fiber-based optical frequency comb sensing cavity. *Opt. Express* **27**, 21463-21476 (2019).

[31]   Oe, R., Minamikawa, T., Taue, S., Nakahara, T., Koresawa, H., Mizuno, T., T., Yamagiwa, M., Mizutani, Y., Yamamoto, H. & Yasui, T. Improvement of dynamic range and repeatability in a refractive-index-sensing optical comb by combining saturable-absorber-mirror mode-locking with an intracavity multimode interference fiber sensor. *Jpn. J. App. Phys.* **58**, 060912 (2019).

[32]   Nakajima, Y., Kusumi, Y. & Minoshima, K. Mechanical sharing dual-comb fiber laser based on an all-polarization-maintaining cavity configuration. *Opt. Lett.* **46**, 5401-5404 (2021).

[33]   Antonio-Lopez, J. E., Castillo-Guzman, A., May-Arrioja, D. A., Selvas-Aguilar, R. & LiKamWa, P. Tunable multimode-interference bandpass fiber filter. *Opt. Lett.* **35**, 324-326 (2010).

[34]   Nakahara, T., Oe, R., Kajisa, T., Taue, S., Minamikawa, T. & Yasui, T. Application of refractive-index-sensing optical frequency comb for biosensing of antigen-antibody reaction. in, Technical Digest of Conference on Lasers and Electro-Optics (CLEO) 2021, STu2A.2 (2021).

[35]   Gauglitz, G. Analytical evaluation of sensor measurements. *Anal. Bioanal. Chem.* **410**, 5-13 (2018).

[36]   Arnaout, R., Lee, R. A., Lee, G. R., Callahan, C., Yen, C. F., Smith, K. P., Arora,

R. & Kirby, J. E. SARS-CoV2 testing: the limit of detection matters. *BioRxiv*, PMC7302192 (2020).

[37] Bong, J.-H., Kim, T.-H., Jung, J., Lee, S. J., Sung, J. S., Lee, C. K., Kang, M.-J., Kim, H. O. & Pyun, J.-C. Pig Sera-derived Anti-SARS-CoV-2 antibodies in surface plasmon resonance biosensors. *Biochip J.* **14**, 358-368 (2020).

[38] Qiu, G., Gai, Z., Tao, Y., Schmitt, J., Kullak-Ublick, G. A. & Wang, J. Dual-functional plasmonic photothermal biosensors for highly accurate severe acute respiratory syndrome coronavirus 2 detection. *ACS Nano* **14**, 5268-5277 (2020).

[39] Moitra, P., Alafeef, M., Dighe, K., Frieman, M. B. & Pan, D. Selective naked-eye detection of SARS-CoV-2 mediated by N gene targeted antisense oligonucleotide capped plasmonic nanoparticles. *ACS Nano* **14**, 7617-7627 (2020).

[40] Zhao, X., Hu, G., Zhao, B., Li, C., Pan, Y., Liu, Y., Yasui, T. & Zheng, Z. Picometer-resolution dual-comb spectroscopy with a free-running fiber laser. *Opt. Express* **24**, 21833-21845 (2016).

[41] Zhao, X., Li, T., Liu, Y., Li, Q. & Zheng, Z. Polarization-multiplexed, dual-comb all-fiber mode-locked laser. *Photonics Res.* **6**, 853-857 (2018).

[42] Nakajima, Y., Hata, Y. & Minoshima, K. High-coherence ultra-broadband bidirectional dual-comb fiber laser. *Opt. Express* **27**, 5931-5944 (2019).

## Acknowledgements

This work was supported by a Japan Society for the Promotion of Science (JSPS) Grant-in-Aid for Scientific Research with grant number 22H00303 and the Japan Agency for Medical Research and Development (AMED) with grant number 20he0822006j00. This work was also supported in part by JST Moonshot R&D grant number JPMJMS2025 (K.Y. and A.S.). We acknowledge the financial support for the project from the Promotion of Regional Industries and Universities by the Cabinet Office, as well as from the Plan for Industry Promotion and Young People's Job Creation by the Creation and Application of Next-Generation Photonics by Tokushima Prefecture. The authors acknowledge Prof. Kaoru Minoshima at The University of Electro-Communications, Japan and Dr. Yoshiaki Nakajima of Toho University, Japan for their help in the mechanical sharing dual-comb configuration.

**Author contributions**

Tak.Yas. and T.K. conceived the project. S.M., R.O., and T.N. developed the dual-comb biosensing system, performed the experiments and analysed the data. T.K., S.T., Y.T. and S.O. made the MMI fibre biosensor. K.O. and A.S. performed the ELISA experiment. K.Y., T.S., T.M., T.K., S.T., and Tak.Yan. discussed the results and commented on the manuscript. S.M. and Tak.Yas. wrote the manuscript. All authors reviewed the manuscript.

**Competing interests**

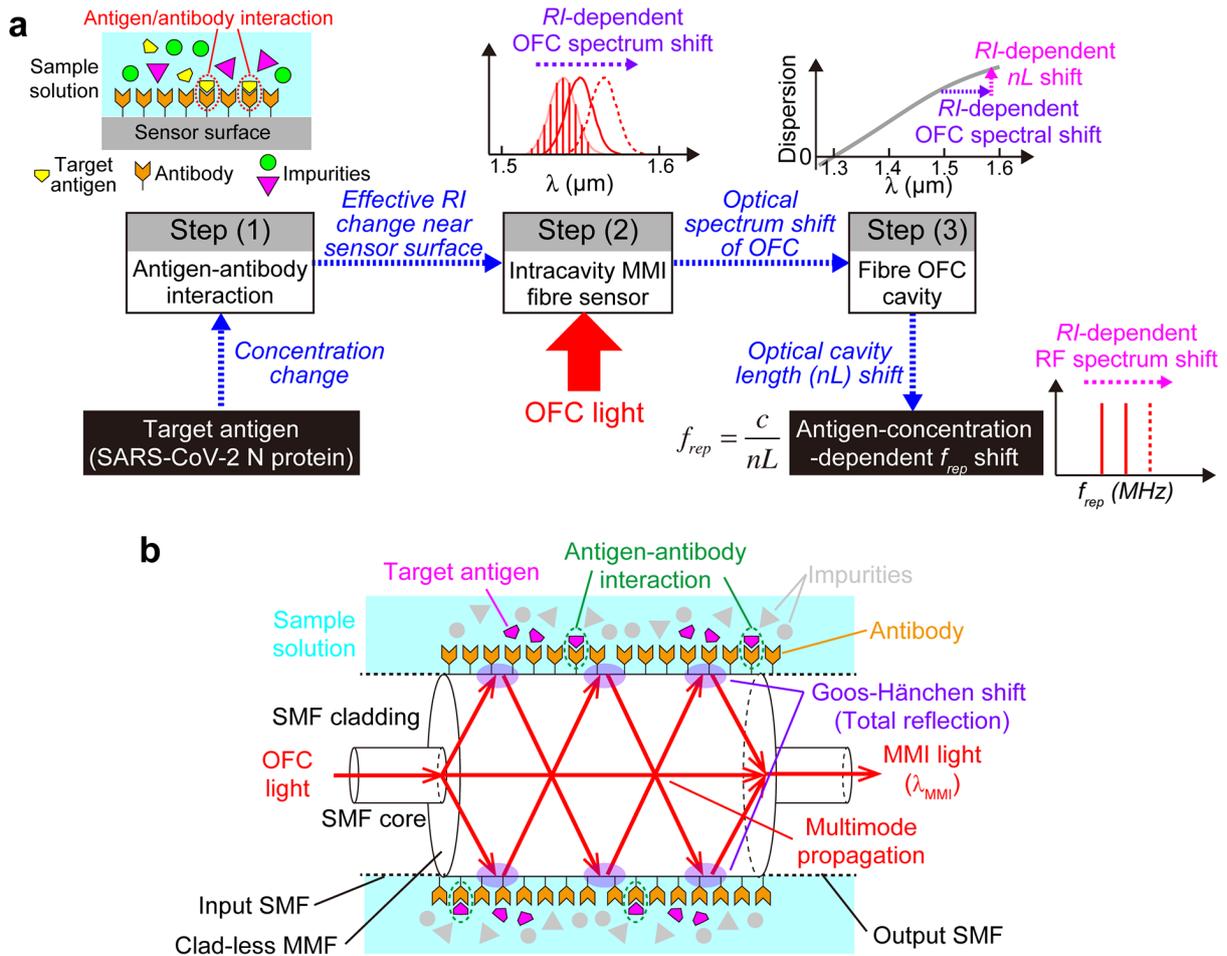

**Figure 1. Principle of operation for the biosensing OFC. a**, Block diagram of the signal flow. The concentration of the target antigen is obtained according to the mode spacing $f_{rep}$ of the OFC through steps (1), (2), and (3). In step (1), the selective combination of a target antigen with the corresponding antibody changes the effective RI near the sensor surface depending on the antigen concentration. In step (2), since the intracavity MMI fibre sensor transmits only certain wavelength ($\lambda_{MMI}$) light based on its RI due to MMI and the Goos-Hänchen shift, the OFC shows an RI-dependent and hence an antigen-concentration-dependent shift in the optical spectrum. In step (3), the antigen-concentration-dependent shift in the optical spectrum is converted to a shift in the optical cavity length $nL$ of the OFC, where $n$ and $L$ are the RI and physical

length of the OFC cavity fibre, via the wavelength dispersion of the cavity fibre. Simultaneously, the intracavity MMI fibre sensor enhances the sensing sensitivity through multiple interactions between the light and sample inside the OFC cavity. **b**, Schematic diagram of the intracavity MMI fibre sensor with antibody surface modification. The surface of the MMI fibre sensor (material = $SiO_2$) was modified with amino-terminated groups through a silane coupling reaction to realize a self-assembled monolayer (SAM) after surface cleaning and modification by UV ozone. Then, the N protein antigen was immobilized on the amino-group-coated MMF fibre sensor surface to realize an active sensing OFC. Additionally, for the dummy sensing OFC, the SAM without immobilized antibody was applied to the surface of the MMI fibre sensor. The combination of a commercialized N protein monoclonal antibody (Fapon Biotech Inc., Dongguan, Guangdong, China, FPZ0553) and a commercialized N protein recombinant antigen (Fapon Biotech Inc., Dongguan, Guangdong, China, FPZ0513) was used for the antigen-antibody interaction for detection of SARS-CoV-2.

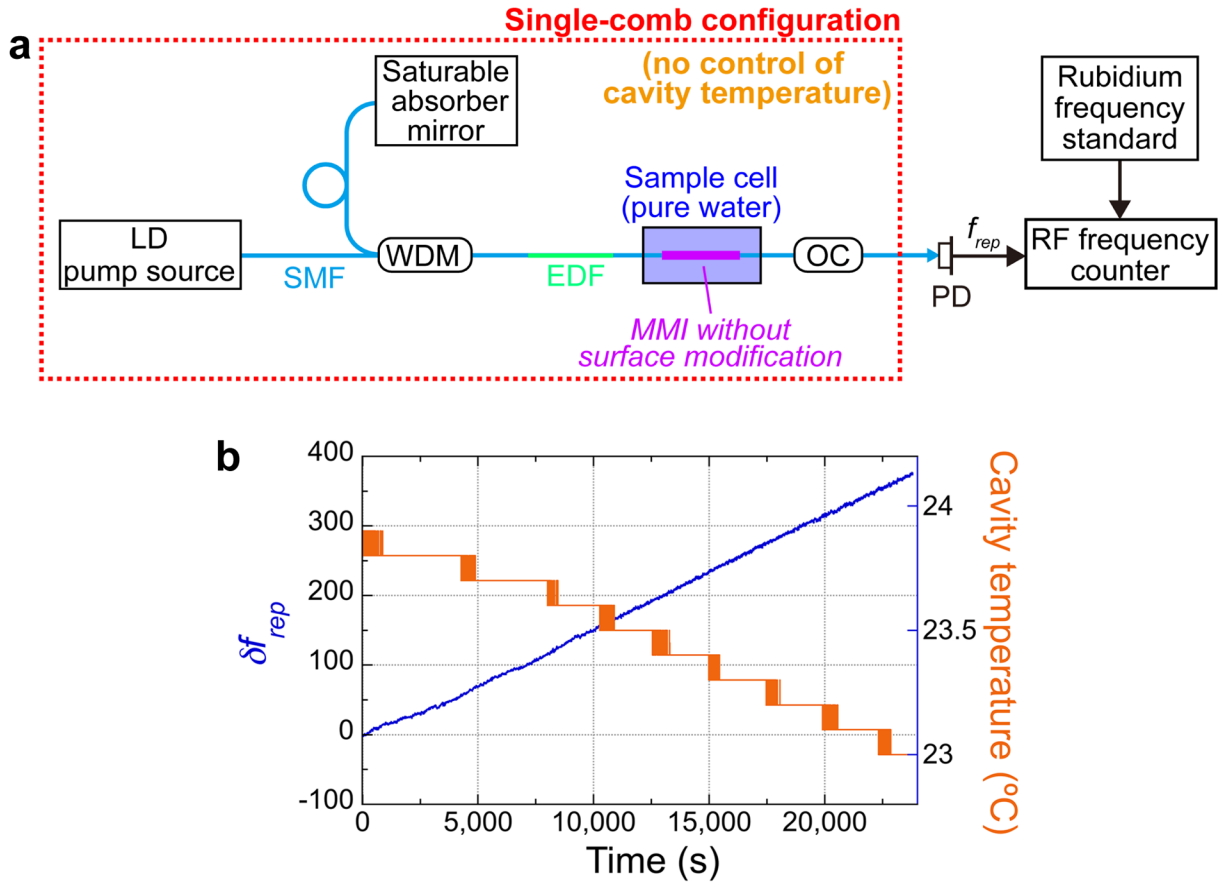

**Figure 2. Basic performance of single-comb RI sensing of pure water with temperature drift. a,** Schematic drawing of the experimental setup. LD, laser diode; SMF, single-mode fibre; WDM, wavelength-division-multiplexing coupler; EDF, erbium-doped fibre; MMI, multimode-interference fibre sensor; OC, fibre output coupler; PD, photodiode. The single sensing OFC has a centre optical wavelength of 1550 nm and a frequency spacing $f_{rep}$ of 32.5 MHz. Details of the single sensing OFC are given in the Methods section. **b,** Temporal fluctuation of the cavity temperature (orange line) and the corresponding $f_{rep}$ shift ($\delta f_{rep}$, blue line). Pure water was used as the sample. $\delta f_{rep}$ was calculated as the frequency deviation from the initial value of $f_{rep}$. The stepped-down behaviour of the cavity temperature is due to the temperature

resolution of the thermistor (= 0.1 °C) used for monitoring the cavity temperature.

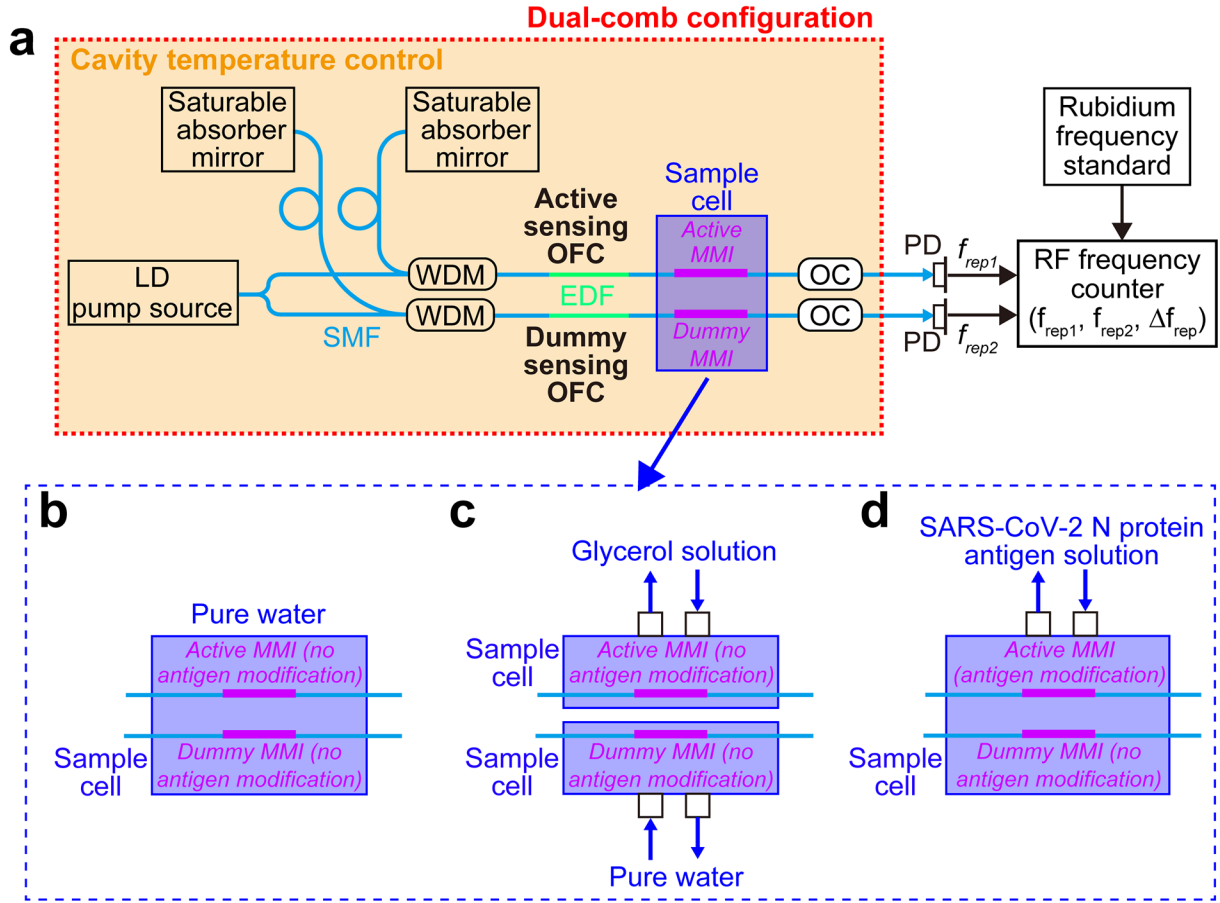

**Figure 3. Experimental setup of dual-comb RI sensing and biosensing. a,** Schematic drawing of the whole experimental setup. LD, laser diode; SMFs, single-mode fibres; OFCs, optical frequency combs; WDMs, wavelength-division-multiplexing couplers; EDF, erbium-doped fibre; active MMI, intracavity multimode-interference fibre sensor for a target sample; dummy MMI, intracavity multimode-interference fibre sensor for a reference sample; OCs, fibre output couplers; PDs, photodiodes. The active sensing OFC and dummy OFC operate at a centre optical wavelength of 1550 nm. **b,** Schematic drawing of the MMI sensor and sample cell for dual-comb RI sensing of pure water. **c,** Schematic drawing of the MMI sensor and sample cell for dual-comb RI sensing of glycerol solution. **d,** Schematic drawing of the MMI sensor and sample cell for dual-comb biosensing of the SARS-CoV-2 N protein antigen. Peristaltic pumps were used for sample exchange in the dual-comb RI sensing of glycerol solution and the dual-comb biosensing of the SARS-CoV-2 N protein antigen.

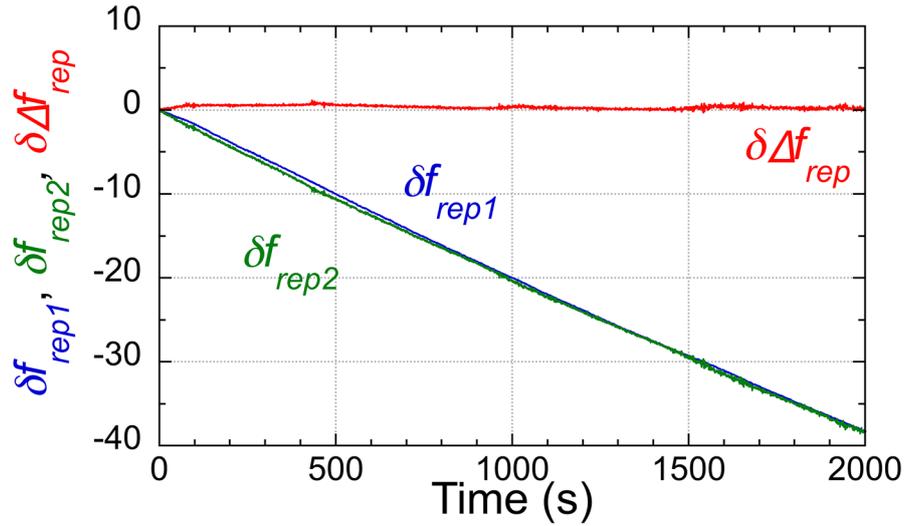

**Figure 4. Basic performance of dual-comb RI sensing of pure water with temperature drift.** Temporal fluctuations in $\delta f_{rep1}$, $\delta f_{rep2}$, and $\delta \Delta f_{rep}$ when pure water was used as a sample for the active and dummy sensing OFCs (see Table 1 and Fig. 3b). $\delta f_{rep1}$, $\delta f_{rep2}$, and $\delta \Delta f_{rep}$ were calculated as the frequency deviations from the initial values of $f_{rep1}$, $f_{rep2}$, and $\Delta f_{rep}$, respectively.

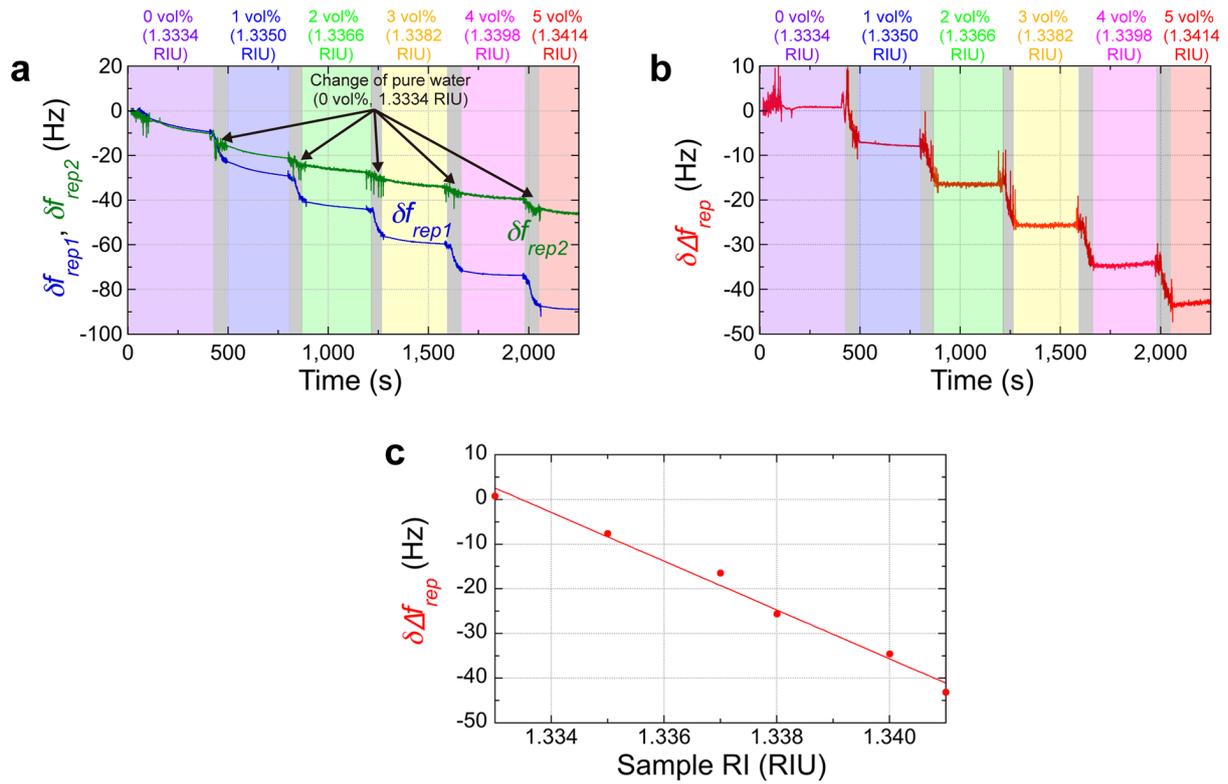

**Figure 5. Temperature-drift-free dual-comb RI sensing of glycerol solution. a**, Sensorgrams of $\delta f_{rep1}$ in the active sensing OFC and $\delta f_{rep2}$ in the dummy sensing OFC. Glycerol solutions consisting of glycerine and pure water at different ratios were used as the target samples in the active sensing OFC; pure water was used as the reference sample in the dummy sensing OFC (see Table 1 and Fig. 3c). Grey zones indicate the time period for sample exchange by peristaltic pumps. $\delta f_{rep1}$ and $\delta f_{rep2}$ were calculated as the frequency deviations from the initial values of $f_{rep1}$ and $f_{rep2}$, respectively. **b**, Sensorgram of $\delta\Delta f_{rep}$ with mixtures of glycerine and pure water at different ratios. $\delta\Delta f_{rep}$ was calculated as the frequency deviation from the initial value of $\Delta f_{rep}$. The mean and the standard deviation of $\delta\Delta f_{rep}$ were 0.76±0.19 Hz at 0 vol%, -7.58±0.24 Hz at 1 vol%, -16.48±0.52 Hz at 2 vol%, -25.64±0.53 Hz at 3 vol%, -34.59±0.31 Hz at 4 vol%, and -43.12±0.34 Hz at 5 vol%. **c**, Relationship between the sample RI and $\delta\Delta f_{rep}$. A linear

relationship between the sample RI and $\delta\Delta f_{rep}$ was obtained with a correlation coefficient (R) of 0.9935.

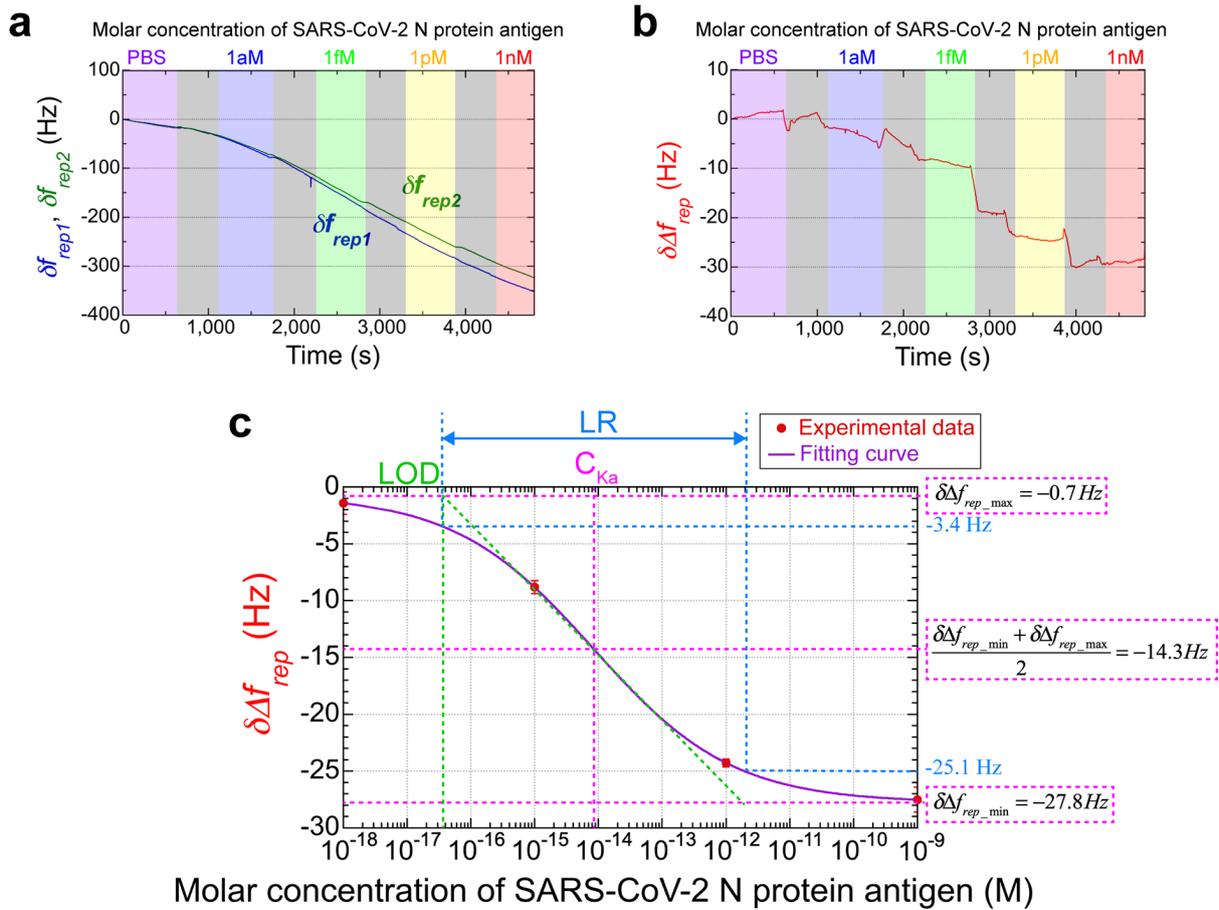

**Figure 6. Rapid, high-sensitivity, dual-comb biosensing of the SARS-CoV-2 N protein antigen. a**, Sensorgram of $\delta f_{rep1}$ and $\delta f_{rep2}$ with respect to different molar concentrations of the SARS-CoV-2 N protein antigen. $\delta f_{rep1}$ and $\delta f_{rep2}$ were calculated as the frequency deviations from the initial values of $f_{rep1}$ and $f_{rep2}$, which were measured for the active biosensing OFC with immobilized antibody and the dummy OFC without immobilized antibody, respectively (see Table 1 and Fig. 3d). **b**, Sensorgram of $\delta\Delta f_{rep}$ with respect to different molar concentrations of the SARS-CoV-2 N protein antigen. $\delta\Delta f_{rep}$ was calculated as the frequency deviation from the initial value of $\Delta f_{rep}$. **c**, Relationship between the antigen molar concentration and $\delta\Delta f_{rep}$. Red dots show experimental data obtained from the sensorgram of $\delta\Delta f_{rep}$. The purple line

shows the fitting curve for the sigmoidal function of the Hill plot, in which coefficient of determination ($R^2$) of 0.9999 was obtained within the range of 1 aM to 1 nM.

**Table 1. Experimental settings of dual-comb RI sensing and biosensing.**

| Experiments | $f_{rep1}$ (MHz) | $f_{rep2}$ (MHz) | $\Delta f_{rep}$ (kHz) | Active MMI | Dummy MMI | Sample cell |
|---|---|---|---|---|---|---|
| Dual-comb RI sensing of pure water | 31.7 | 32.5 | -851.9 | No surface modification | | Single (Fig. 3b) |
| Dual-comb RI sensing of glycerol solution | 31.7 | 32.5 | -851.9 | No surface modification | | Dual (Fig. 3c) |
| Dual-comb biosensing of SARS-CoV-2 N protein antigen | 29.6 | 29.7 | -88.6 | Surface modification of antibody | No surface modification of antibody | Single (Fig. 3d) |

### Table 2. Comparison of testing methods for SARS-CoV-2

| Method | Limit of detection (LOD) | Linear range (LR) | Time | Ref. |
|---|---|---|---|---|
| RT-PCR | 0.17 aM | - | 4~5 hours | [36] |
| SPR | 1 pM | 2~1000 pM | - | [37] |
| Nanoplasmonic-enhanced SPR | 85 fM | 85 fM ~ 2 pM | 5 min | [14] |
| Dual-functional plasmonic biosensor | 220 fM | 10 fM ~ 50 µM | - | [38] |
| Colorimetric assay | 18 pM | 20~300 pM | 10 min | [39] |
| Dual-comb biosensing | 38 aM | 34 aM ~ 2.1 pM | 10 min | - |